# Statistical characterization of tissue images for detection and classification of cervical precancers

*Jaidip Jagtap[1], Nishigandha Patil[2], Chayanika Kala[3], Kiran Pandey[3], Asha Agarwal[3] and Asima Pradhan[1, 2*]*

[1]Department of Physics, IIT Kanpur, U.P 208016
[2]Centre for Laser Technology, IIT Kanpur, U.P 208016
[3]G.S.V.M. Medical College, Kanpur, U.P. 208016

* Corresponding author: asima@iitk.ac.in, Phone: +91 512 259 7971, Fax: +91-512-259 0914.

**Abstract**

Microscopic images from the biopsy samples of cervical cancer, the current "gold standard" for histopathology analysis, are found to be segregated into differing classes in their correlation properties. Correlation domains clearly indicate increasing cellular clustering in different grades of pre-cancer as compared to their normal counterparts. This trend manifests in the probabilities of pixel value distribution of the corresponding tissue images. Gradual changes in epithelium cell density are reflected well through the physically realizable extinction coefficients. Robust statistical parameters in the form of moments, characterizing these distributions are shown to unambiguously distinguish tissue types. These parameters can effectively improve the diagnosis and classify quantitatively normal and the precancerous tissue sections with a very high degree of sensitivity and specificity.

## 1. Introduction

Cancer is a leading cause of death worldwide, with cervical cancer being the fifth most common cancer in women [1-2]. It originates as a few abnormal cells in the initial stage and then spreads rapidly. Treatment of cancer is often ineffective in the later stages, which makes early detection the key to survival. Pre-cancerous cells can sometimes take 10-15 years to develop into cancer and regular tests such as pap-smear are recommended. However, the reliability of the test is not perfect. During the progress of pre-cancers, the morphological and architectural changes involved are, increase in nuclear size, nucleus to cytoplasmic ratio, change in refractive index, hyperchromatism and pleomorphism [3-5]. These changes are observed under a microscope and provide an opportunity for physicians to offer diagnosis and treatments. However, distinguishing the normal from abnormal or suspicious tissue is a non-trivial task.

Conventional techniques such as MRI, tomography, digital mammography, ultrasound and colposcopy often fall short in terms of specificity or resolution or are time-consuming [6]. The current “gold standard” for detecting cancer of the epithelial tissue is the histopathology analysis of biopsy samples seen under the microscope. Samples are removed from the patient before being sectioned, fixed, stained, and examined by a pathologist for morphological abnormalities. In connection with cervical pre-cancers, the various stages display morphological changes along the epithelium layer. The pathologist observes the nucleus to cytoplasm ratio and the density of cells in the epithelium layer to classify the different grades of dysplasia. This technique is subjective since it relies heavily on the pathologist’s qualitative analysis of the microscopic image.

Quantification of the histopathological data is a solution to such problems. Hence several studies [7-9] have been carried out to develop image analysis tools which can classify the different grades of dysplasia in human cervical tissue and reduce the detection error made by the pathologists' subjective inference. Most of the present research in image analysis is based on image segmentation as a potential means of discriminating between normal and abnormal tissue sections [7, 9]. Reddy et al [9] used image segmentation techniques on digital mammograms for classification which shows better visualization. Reliable classification depends on number of well separated classes in the image and effective noise removing filters. Collier et al [10] have used a segmentation algorithm on cervical tissue images obtained using confocal microscopy for locating and isolating nuclei in the image. They then estimated scattering coefficient in different regions of normal and precancerous cervical epithelium. Gossage et al [7] have performed texture analysis of OCT images for mouse skin, fat, normal lung, and abnormal lung in in-vivo tissue classification through histogram characterization. The extracted parameters are used as additional features in classification algorithms involving Neural networks [11].

We present here a statistically robust method which may provide additional useful information for classification. Keeping in mind the crucial role of microscopic images for histopathological tissue characterization, we carry out a systematic investigation of the statistical features of the normal and precancerous cervical tissue images after clearly establishing differing correlation properties in normal and different grades of pre-cancer. Parameters such as mean, variance and more informative higher moments like skewness and kurtosis well capture the differing properties of tissues. It is worth mentioning that lower order moments have been considered by Bountris et al 8 for early stage detection of bronchial cancer. They used white and blue light images for texture analysis. Their study indicates that techniques like first order statistical analysis performed effectively on white light bronchoscopy images for diagnosing abnormal regions. Maussang et al [12] have also used skewness and kurtosis for classification of Synthetic Aperture Sonar (SAS) imagery data. To provide a better physical sense, images of extinction coefficients are calculated from the microscopic images and show a very good correlation with cell density changes along the depth of the epithelium.

Our study on microscopic images shows a promising approach for early stage detection and quantitative classification of precancerous grades from biopsy images which will supplement the pathologists' qualitative analysis.

## 2. Experimental Setup

The pathologically graded (CIN or dysplastic) biopsy samples of human cervical tissue are illuminated by an integrated halogen lamp controlled by an SMPS (6V-20W) and imaged under a simple microscope (LABOMED VISION 2000) with 20X objective and digital camera (Sony, DSC-W110, 7.2MP). The tissue slides contain 5μm thick transverse tissue sections having lateral dimensions of 4mm×6mm. The normal counterparts are obtained from the region adjacent to the abnormal region of the resected tissue. For preparation of the sections, the tissue is first dehydrated and embedded in wax. Subsequent sectioning is undertaken with the help of a rotary microtome [13]. It may be noted that the tissue sections used in this study are unstained. The slides are characterized from histopathology reports. The validation set is compared with the same after image analysis, for determining specificity of the system. 53 tissue sections were collected from G.S.V.M. Medical College and Hospital, Kanpur, India from patients in the age group of 35−60 years. All the cervical tissues included in this study are of the stratified squamous epithelium type.

The images are captured at an optical zoom of 4X. Since the epithelium region thickness varies according to the patient’s age and location, the captured images are cropped as per the epithelium layer thickness before image segmentation. Thus the image sizes vary from 500 X 150 pixels to 800X 350 pixels, restricting the analysis to the epithelial layer.

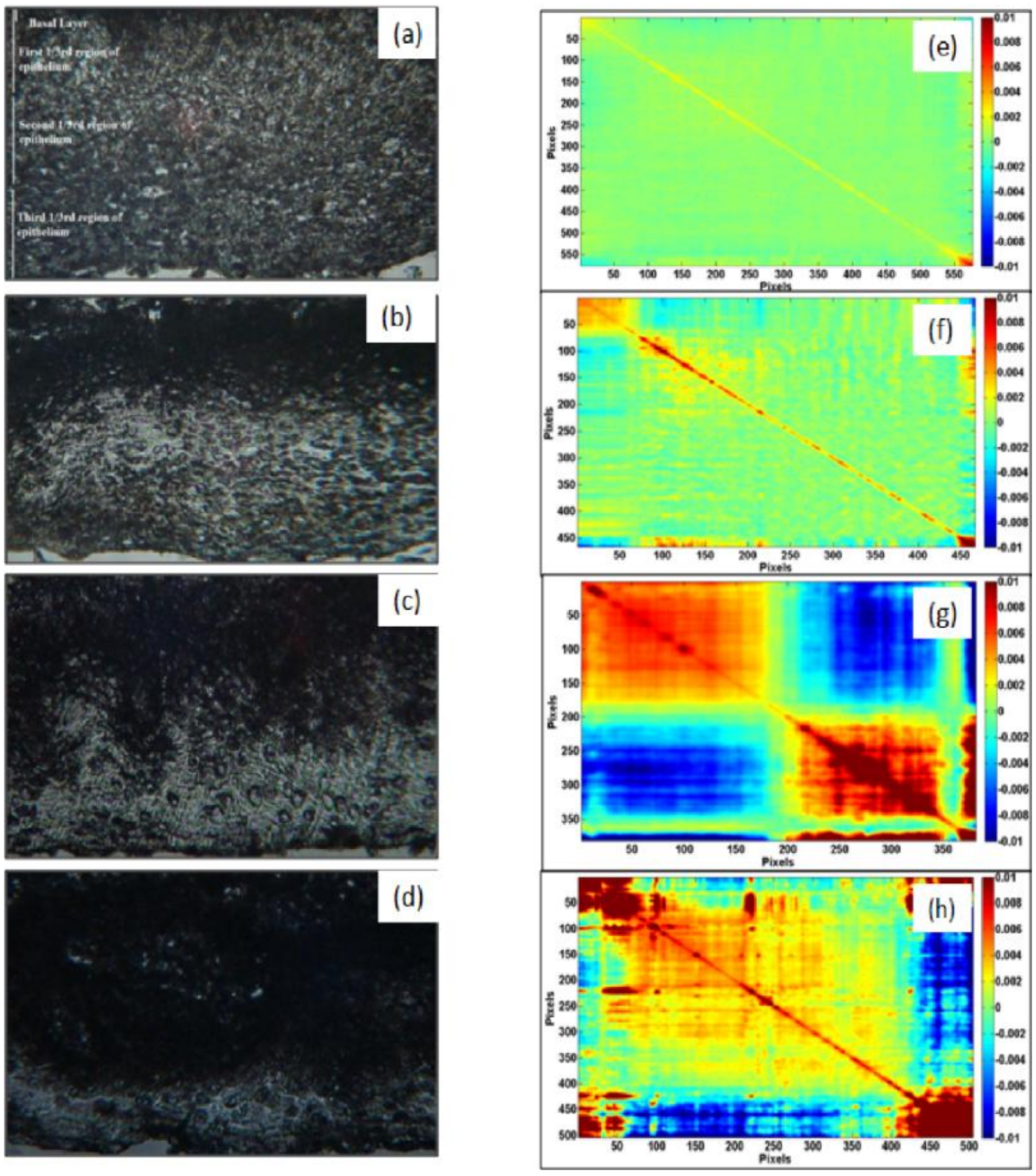

**Figure 1** Typical microscopic images and corresponding correlation maps (color online) of **a)** & **e)** normal, **b)** & **f)** Gd-1, **c)** & **g)** Gd-2, **d)** & **h)** Gd-3 of human cervical tissue sections respectively. The correlation matrix images clearly indicate differing correlation behavior with progress of the grades of precancer.

Typical microscopic images for normal and various grades of dysplasia (Gd) are as shown in figure 1 (a-d). The images represent the epithelium, starting from basal layer at the top of the image. Increase in cell density along the epithelium can be seen as the grade of dysplasia increases. A high cell density is noticed in the first $1/3^{rd}$ of the epithelium in Gd1 dysplasia as seen in figure 1(b). In Gd2 the cell density is high upto $2/3^{rd}$ of the epithelium (figure 1(c)) while a uniformly high cell density in the entire epithelium is noticed in Gd3. The pathologists use this criterion to classify the different grades.

**3. Method**

The microscopic images display a variation in the cell density along the epithelium at different stages of pre-cancer. These images are subjected to auto correlation analysis. After clearly establishing differing correlation properties in normal and different grades of pre-cancer a systematic investigation of the statistical features of the normal and precancerous tissue images is carried out. The fact that correlation behavior indicates increasing cell clustering as the disease progresses clearly manifests in the respective pixel probabilities and their statistical parameters. As mentioned earlier, digital images of the tissue slides are cropped prior to processing using an image editing application (Microsoft Windows Picture Manager) to restrict the analysis area to the epithelium layer. Since the size thickness of the epithelium varies for different patients, the images are not uniform in terms of pixel size. Hence all the data have to be normalized with respect to the image size in each case for comparison between different images. Normalization may be done either on the basis of histogram maxima or total area under consideration. Since this method relies on statistical analysis, we normalized on the basis of area, as the maxima values can be a statistical parameter. The digital images are stored in jpg format and hence the data is contained in a 5 dimensional array of the form M×N×R×G×B where M×N gives the total number of pixels in the image and the dimensions corresponding to R, G and B store the RGB information of the image.

*3.1 Correlation Matrix Images:*

In order to identify the structurally similar areas, correlation matrix has been computed as $C=A^TA/N$ where A is the mean subtracted microscope image matrix normalized by the standard deviation at each pixel. The correlation matrix delineates the image into domains of structurally similar characteristics. Figure 1 (e-h) shows typical correlation matrix images for normal and various grades of dysplasia in human cervical tissue sections. The significant differences seen in the correlation maps of normal, dysplasia of grades 1, 2 and 3 indicate high correlations at high density regions which correspond to increased clustering of cells. This indicates the statistical properties of the pixel value distribution which can show significant difference among tissue types.

*3.2 Pixel probability distribution and characterization of their moments:*

Subsequent to observations of the differing correlation properties in different tissue types, various statistical moments are analyzed for a finer analysis. For studying the characters of the intensity distributions, the RGB images are converted to a grayscale image in MATLAB using built-in functions. Typical area normalized histograms for normal tissue and grade1-3 images are shown in figure 2. The grayscale level of intensity values ranges from 0 to 255, where 0, corresponds to black and 255, to white. The area normalized histogram data thus give us information of the number of the pixels corresponding to each grayscale level. As can be observed from figure 2, the histograms with different grades of dysplasia and normal tissue are considerably different. The peak value, FWHM and the various statistical moments such as mean, standard deviation, skewness and kurtosis are then applied to quantify these differences. For the standard definitions of the statistical moments, one may refer to Bountris et al [8].

The problem at hand is now considered to be of multimodal classification by using the various statistical moments. The aim is to determine if multiple parameters can be used for classification. The receiver operating characteristic (ROC) [14-15] analysis is first performed on training data of 30 tissue images to determine the sensitivity and specificity of each parameter for determining thresholds for finer classification. 23 tissue images are kept aside as a validation set. These are first classified based on each parameter individually on the basis of the thresholds fixed by the ROC analysis for the training set. Multiple parameters are then considered simultaneously and the validation set is reclassified.

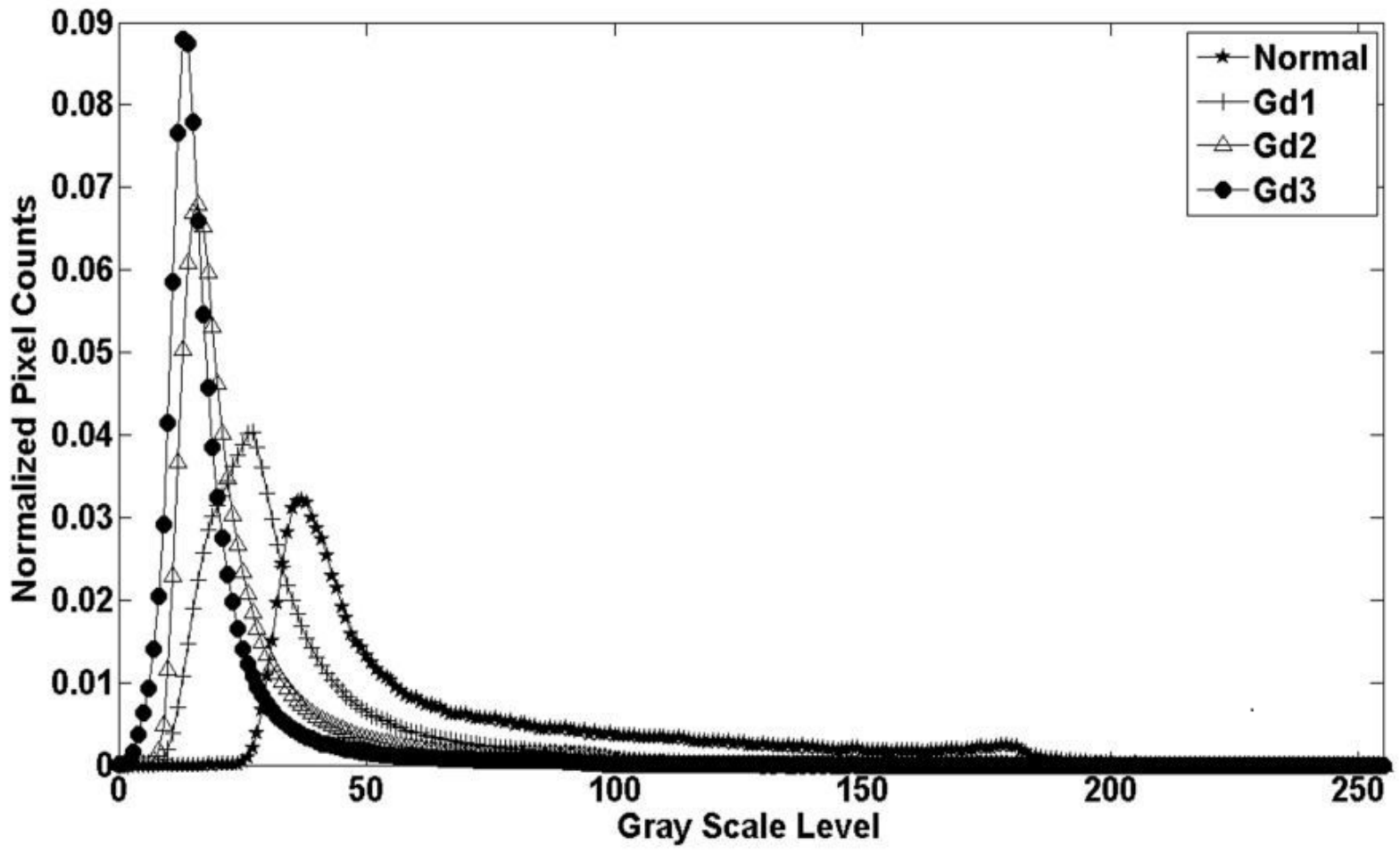

**Figure 2** Typical area normalized histograms for normal and dysplastic tissue of various grades.

*3.3 Extinction Coefficient:*

The basis of optical diagnosis of cervical precancers is the increase in cell density and consequently a change in the optical parameters, specifically the scattering coefficient. Light propagation in thin tissue sections obeys Beer-Lambert law [16]. Progression of precancers to different stages is histopathologically evaluated based on the depth up to which the cell density has increased in the epithelium [13]. Correspondingly, the extinction coefficients would change with depth in the epithelium. For a physical realization it is sometimes necessary to use the optical parameters for discrimination. An image of extinction coefficient may provide a better contrast than just intensity based images. To examine this, the microscopic images are further used to determine the extinction coefficient (scattering coefficient + absorption coefficient). If $I_0$ is intensity of the source, incident on a thin tissue section of thickness Z, then the emerging intensity, $I(z)$, will follow the Beer-Lambert law [16] as

$$I(z) = (1 - R_F) I_0 \exp(-\mu_z z) \qquad (2)$$

where extinction coefficient $\mu_z = \mu_s + \mu_a$ ; $\mu_s$ is scattering coefficient and $\mu_a$ is absorption coefficient of the tissue. In typical tissues the absolute values of absorption coefficient may vary from $10^{-2}$ to $10^{4}$ $cm^{-1}$. [16] $R_F$ is the Fresnel reflection coefficient and at normal beam incidence $R_F = [(n-1)/(n+1)]^2$ where n =1.4 is the relative refractive index of tissue and surrounding media. $I_0$ is measured by allowing the incident light to pass through only the glass slide. Similarly an image captured when light travel through a tissue section placed on the same glass slide measures $I(z)$. The extinction coefficients at every pixel is calculated for a 5μm thick tissue sections of normal and various grades of cervical dysplasia. The average of all extinction coefficient pixels is calculated as the extinction coefficient of each sample.

## 4. Results and Discussion

Table 1 presents the mean values of the statistical parameters for each tissue type. We observe that while it is easy to distinguish precancerous tissue from normal tissue based on the parameters in table 1, the difference between the grades themselves is relatively smaller as seen from the individual values and hence these parameters by themselves may seem insufficient. The normalized pixel counts were plotted for various grayscale levels between 15 and 40, chosen based on the grayscale level corresponding to maxima across the four classes.

The weighted mean decreases with increase in density of cells as expected since it is a measure of the average intensity transmitted. A difference in cell density along the length of the epithelium appears more in the grades of dysplasia and this is quantitatively represented by the standard deviation. A more random cell density in normal epithelium leads to a higher standard deviation.

**Table 1** Mean values of statistical parameters across the various grades.

| PARAMETER | Normal | Grade 1 | Grade 2 | Grade 3 |
|---|---|---|---|---|
| Weighted mean | 64.45 | 33.62 | 25.29 | 18.78 |
| Standard deviation | 34.75 | 18.45 | 15.88 | 12.29 |
| Skewness | 2.23 | 2.78 | 4.11 | 4.66 |
| Kurtosis | 7.80 | 10.50 | 20.56 | 25.74 |
| Full width half maxima | 26 | 17 | 9.9 | 8.5 |
| Maxima (FWHM) | 0.03 | 0.04 | 0.07 | 0.10 |
| Grayscale level corresponding to maxima | 38.83 | 20.17 | 16.83 | 13.25 |

The fact that the cell density is not uniform in the higher grades is manifested in the values of skewness which is a measure of the asymmetry of a distribution. Kurtosis, on the other hand, quantifies the sharpness of a distribution which is higher for the higher grades with sharper cell density distributions. Table 2 shows the ROC cut-off value obtained from a 30 sample training data set. Using these parameters the validation data set of 23 samples were classified blindly and the results were compared with histopathology report to determine sensitivity and specificity for various grades.

**Table 2.** ROC cut-off values for class discrimination.

| Statistical Parameters | Normal &Gd-1 | Gd-1&Gd-2 | Gd-2& Gd-3 |
|---|---|---|---|
| Normalized pixel counts corresponding upto gray | 0.01 | 0.45 | 0.75 |
| Normalized pixel counts corresponding upto gray | **0.21** | **0.57** | **0.76** |
| Skewness | **2.57** | **3.37** | **4.92** |
| Kurtosis | **9.33** | **14.51** | **27.51** |
| Entropy | **6.35** | **5.74** | **5.20** |
| Standard Deviation | 24.01 | 17.08 | 14.52 |

.

Table 3. Specificity and sensitivity for various statistical parameters.

| Statistical Parameters | Data | Normal & Gd-1 | | Gd-1 & Gd-2 | | Gd-2 & Gd-3 | |
|---|---|---|---|---|---|---|---|
| | | Specificity (N) | Sensitivity (G11) | Specificity (G12) | Sensitivity (G21) | Specificity (G22) | Sensitivity (G3) |
| Normalized pixel counts corresponding upto gray scale level 22 (P22) | Training Set | 1 | 1 | 1 | 1 | 1 | 1 |
| | Validation Set | 0.66 | 1 | 0.66 | 0.82 | 0.82 | -- |
| | Overall | 0.86 | 1 | 0.86 | 0.92 | 0.92 | 1 |
| Normalized pixel counts corresponding upto gray scale level | Training Set | 1 | 1 | 1 | 1 | 1 | 1 |
| | Validation Set | 1 | 1 | 0.66 | 0.76 | 0.88 | -- |
| 25 (P25) | Overall | 1 | 1 | 0.86 | 0.90 | 0.95 | 1 |
| Statistical Parameters | Data | Normal & Gd-1 | | Gd-1 & Gd-2 | | Gd-2 & Gd-3 | |
| Skewness | Training Set | 1 | 1 | 1 | 1 | 0.95 | 1 |
| | Validation Set | 0.66 | 1 | 0.66 | 0.88 | 0.65 | -- |
| | Overall | 0.86 | 1 | 0.86 | 0.95 | 0.81 | 1 |
| Kurtosis | Training Set | 1 | 1 | 1 | 1 | 0.95 | 1 |
| | Validation Set | 0.66 | 1 | 0.66 | 0.88 | 0.76 | -- |
| | Overall | 0.86 | 1 | 0.86 | 0.95 | 0.86 | 1 |
| Entropy | Training Set | 1 | 1 | 0.75 | 0.90 | 0.83 | 1 |
| | Validation Set | 0.66 | 1 | 0.66 | 0.72 | 0.88 | -- |
| | Overall | 0.86 | 1 | 0.71 | 0.81 | 0.86 | 1 |
| Standard Deviation | Training Set | 1 | 1 | 0.75 | 0.8 | 0.74 | 1 |
| | Validation Set | 1 | 0.66 | 1 | 0.24 | 0.88 | -- |
| | Overall | 1 | 0.86 | 0.86 | 0.53 | 0.80 | 1 |

For each classification the first and second columns specify sensitivity and specificity which are found to be the same for grayscale level 22 and 25. The plots for normalized pixel counts at grayscale level 22 and 25 versus the grayscale level corresponding to maxima are presented in figures 3(a) and 3(b). The differences between grades 2 and 3 and grades 1 and 2 are more distinct for grayscale level 25 as compared to 22. Hence a preference on grayscale level was placed on 25 over 22.

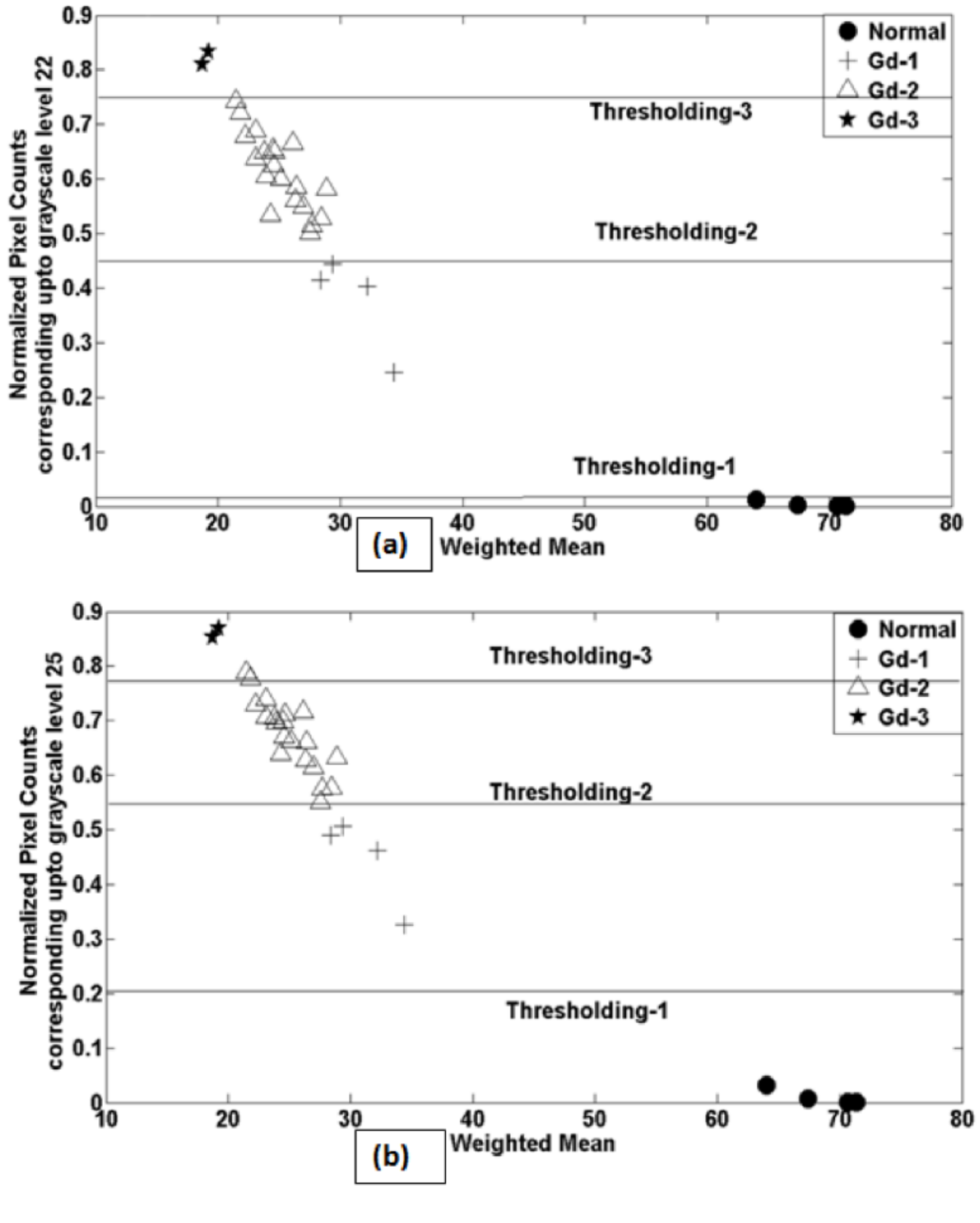

**Figure 3.** Plot of (**a**) normalized pixel counts at grayscale level 22 and (**b**) normalized pixel counts at grayscale level 25 v/s grayscale level corresponding to maxima.

The cell density increases as dysplasia progresses. This is manifested as darker images. Hence lower values of weighted mean are associated with higher grade of dysplasia. The weighted means are considerably distinct; however the standard deviation is almost of the same order among the grades (figure 4). The same also applies for entropy (figure 5).

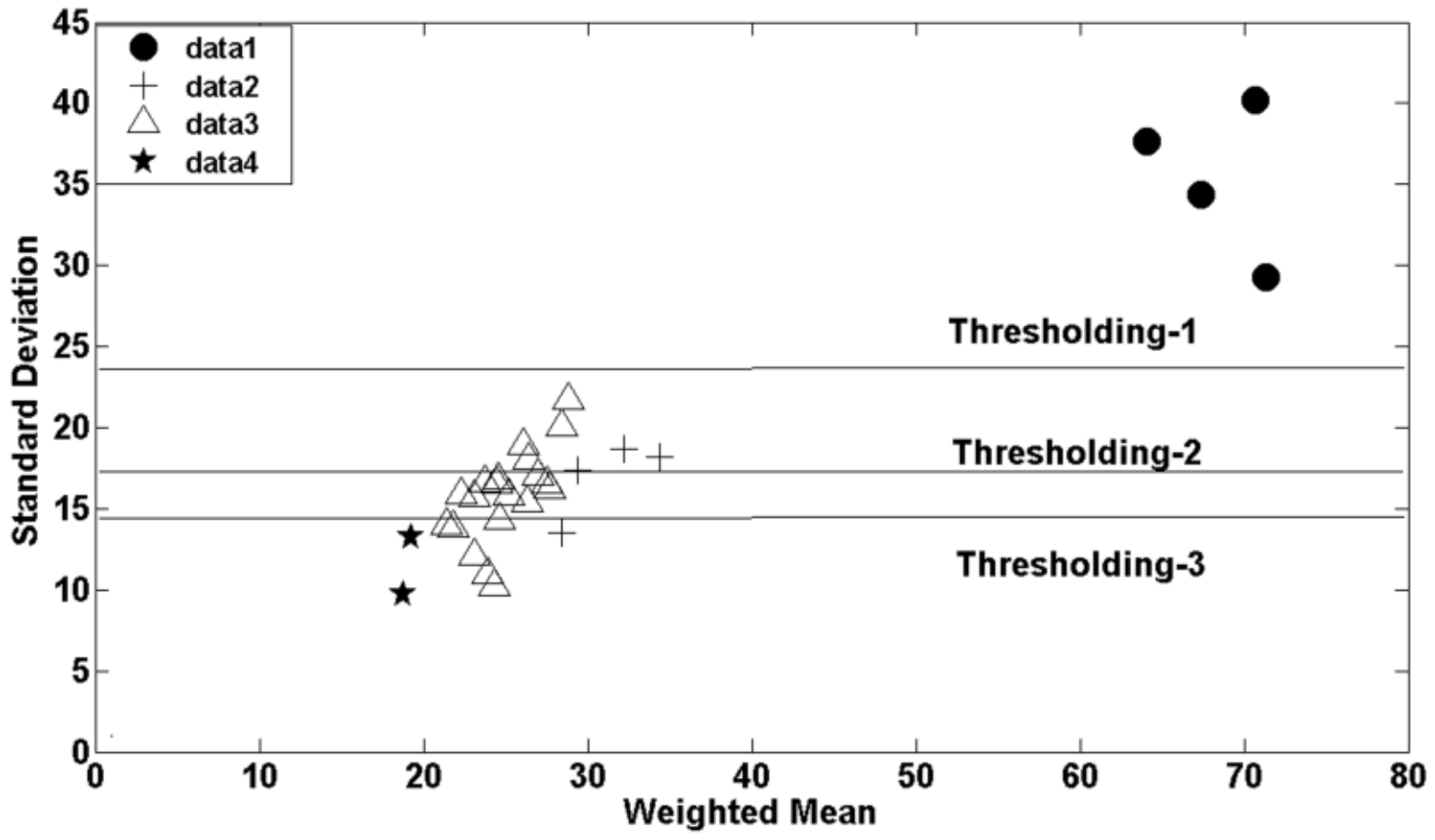

**Figure 4** Plot of standard deviation against weighted mean. Distinct separation is obtained between normal and dysplastic tissue.

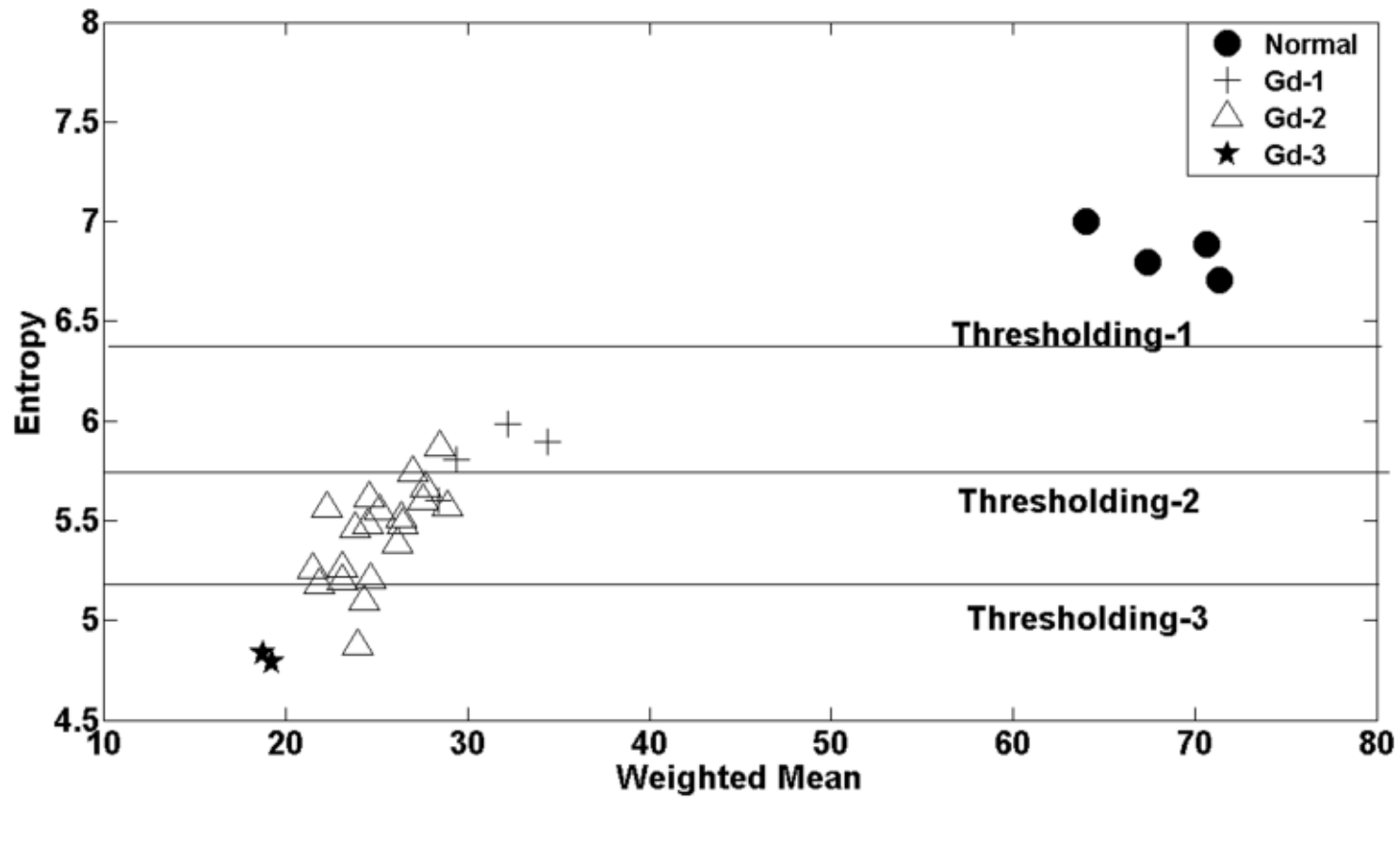

**Figure 5** Plot of entropy against weighted mean results are similar to those obtained with standard deviation (Figure 4).

As mentioned earlier standard deviation is a measure of average contrast i.e. the difference in brightness between different objects in the same field of view. In case of normal tissue, the epithelium has a sparse distribution of cells and the cells are spaced by considerable interstitial substances, thus offering a high contrast. In the dysplastic tissue, the cells are tightly packed and hence a lower value of contrast is obtained. As dysplasia progresses, cell density increases progressively in the epithelial region and often advances into the stromal region. However, there is not much difference in the packing of cells over a region. Thus the average contrast across the grades is of comparable magnitude. A clear distinction is therefore not obtained among the grades by considering weighted mean and standard deviation. However, the higher moments do provide a better classification as seen in figure 6(a) and (b).

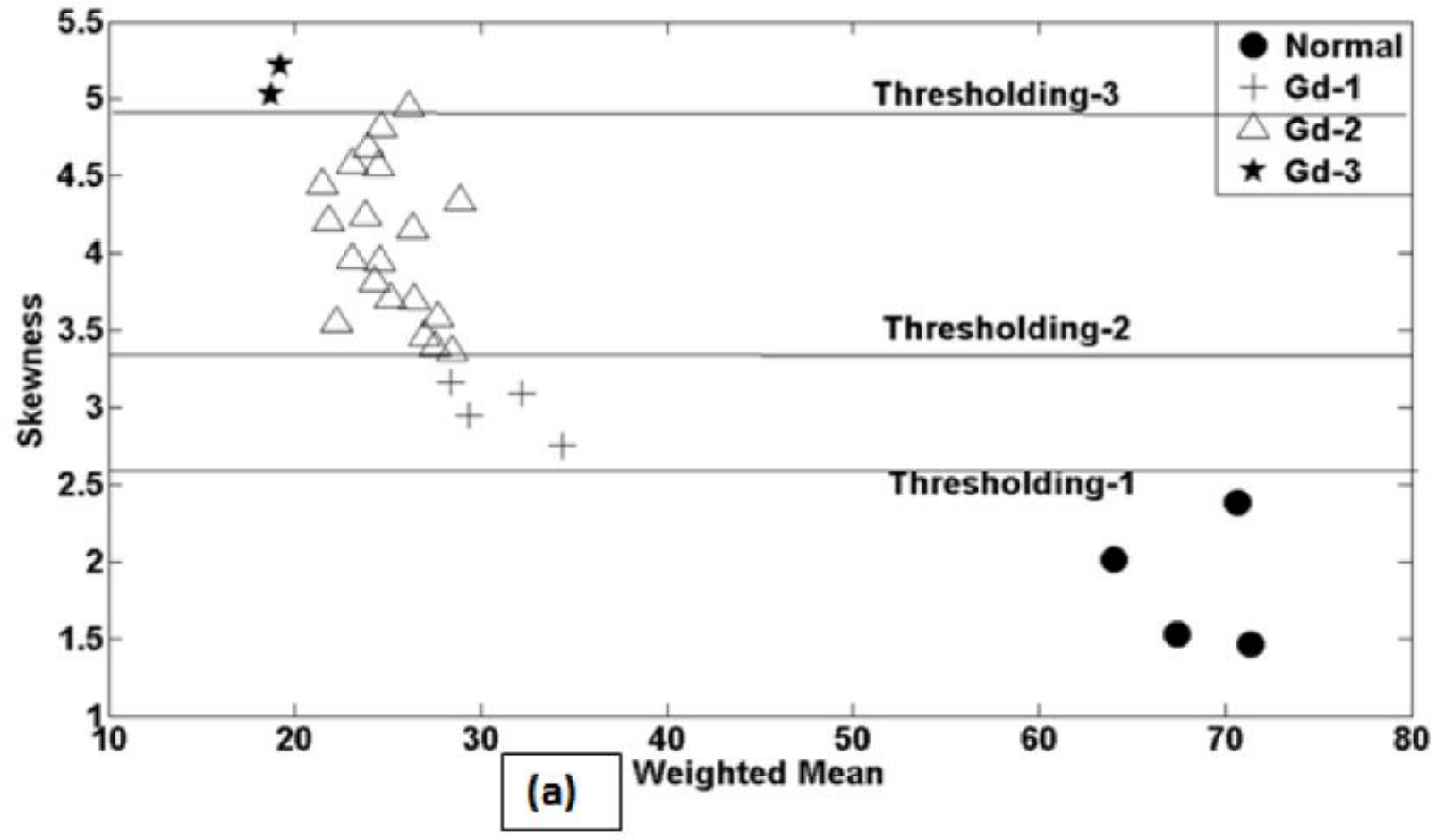

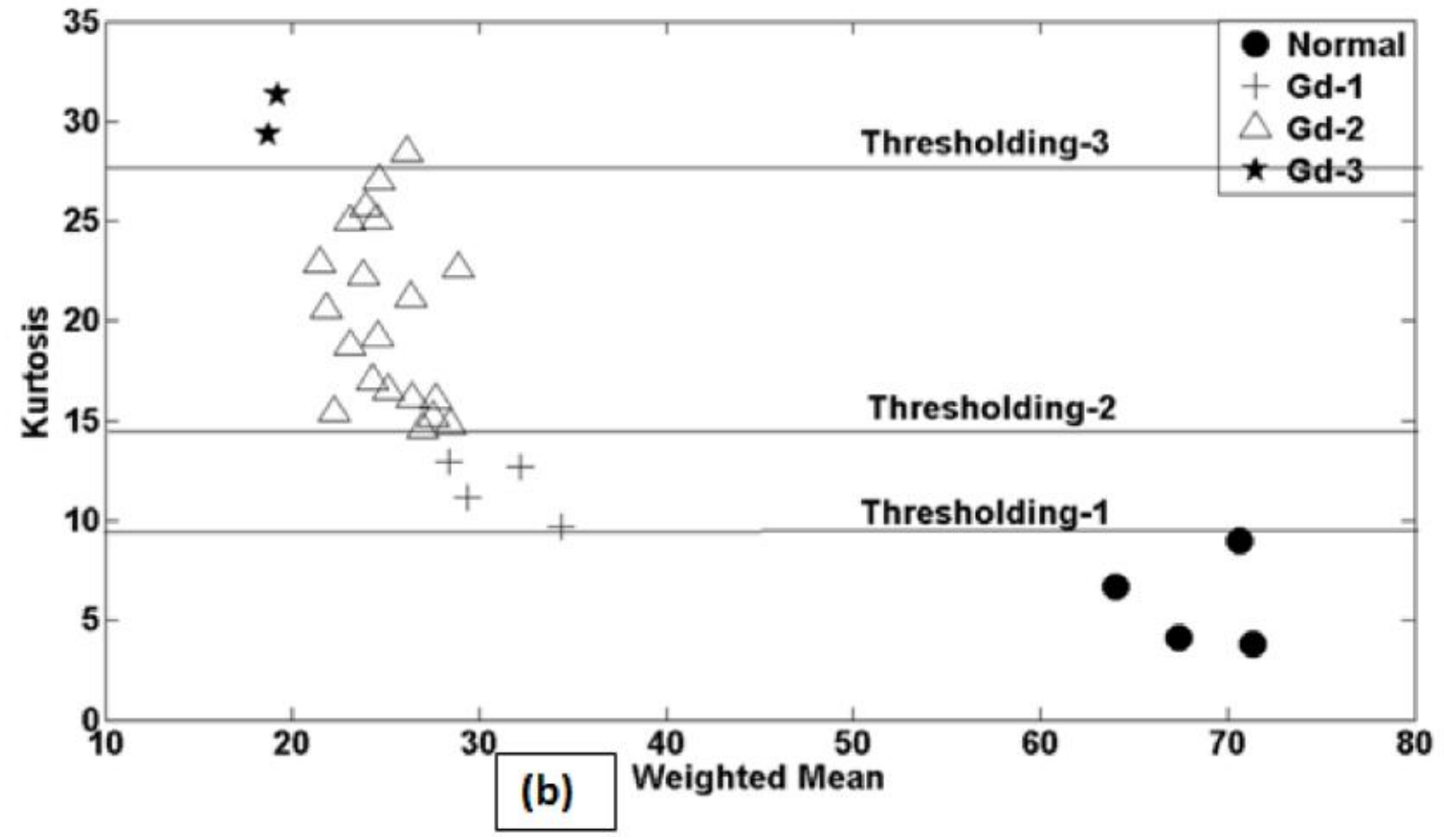

**Figure 6** Shows (**a**) skewness and (**b**) kurtosis plotted against weighted mean.

The sensitivity and specificity for the rest of the parameters such as FWHM, weighted mean and pixel count maxima are found to be too poor to use them for classification. The validation set is then classified on the basis of thresholds fixed by ROC analysis of training set. A classification accuracy of 80% is obtained with grayscale level 22; however, the same improved to 92% using grayscale level 25.

*4.1 Classification Algorithm:*

Further to make the classification sturdy, we use the various parameters in tandem with weights assigned to each in a classification algorithm which is effective for discrimination within the various grades. The weights used in the classifier program are derived from sensitivity and specificity values of training data shown in the Table 3. Sensitivity specifies the true positive rate at a threshold, while the specificity specifies the true negative rate. For example at the first threshold, the true positive rate is the number of grade 1 samples correctly classified, and the true negative rate is the number of normal samples correctly classified. Table 4 shows the statistical parameter weights (last column) calculated from Table 3. The second column in the table is simply the specificity at the first threshold. The third column is the average of the sensitivity at first threshold (G11), and the specificity at the second threshold (G12). Both the thresholds need to be considered here as the number of samples correctly classified in a particular grade is defined by the true positive rate at the lower threshold and the true negative rate at the upper threshold. The fourth column similarly is the average of the true sensitivity at the second threshold (G12) and the specificity at the third threshold (G22) while the fifth column is merely the sensitivity at the third threshold (there being no upper threshold). The last column presents the weights assigned to each parameter in the classifier and are evaluated as an average of the previous 4 columns for each row.

**Table 4** Statistical parameter weight calculation from table 3.

| | Normal(N) | Gd1(=(G11+G12)/2) | Gd2(=(G12+G22)/2) | Gd3 | Weights (=(N+Gd1+Gd2+Gd3)/4) |
|---|---|---|---|---|---|
| P25 | 1 | 1 | 1 | 1 | 1 |
| Skewness | 1 | 1 | 0.975 | 1 | 0.99 |
| Kurtosis | 1 | 1 | 0.975 | 1 | 0.99 |
| Entropy | 1 | 0.875 | 0.865 | 1 | 0.94 |
| std | 1 | 0.875 | 0.77 | 1 | 0.91 |

It is observed that skewness and kurtosis have the highest sensitivity and specificity and so either of them may be used as a parameter for discriminating the normal and various grades of cervical dysplasia from one another. Entropy shows better statistics compared to standard deviation. Thus the grade discrimination is best obtained by exploiting the physical sense of an optimum number of statistical parameters. The optimum sensitivity and specificity of normal and all 3 grades are now obtained by the following form:

$$g= (P25*1) + (S*0.99) + (E*0.94) \qquad (2)$$

where P25 is normalized pixel counts corresponding upto gray scale level 25, S=Skewness and E=Entropy.

Using this form, we arrive at a sensitivity of 98% and specificity of 99% for the validation set.

*4.2 Extinction Coefficient:*

The extinction coefficients calculated at all pixels form an image. These images correspond to microscopic images (figure 1(a-d)) of 5 micron thick tissue sections shown in figure 7 (a-d).

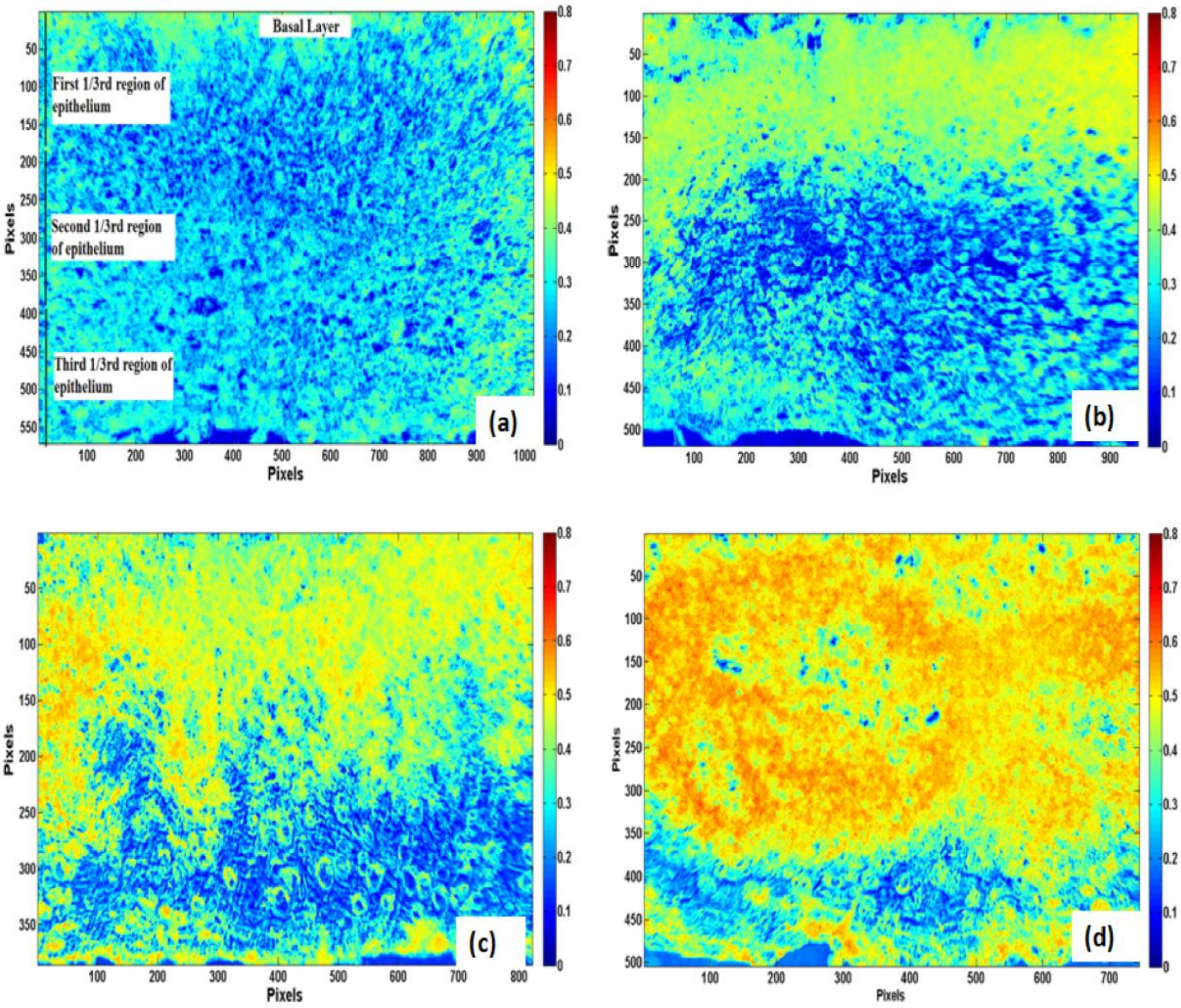

**Figure 7** Extinction coefficient images (color online) of **a**) normal, **b**) Gd-1, **c**) Gd-2, **d**) Gd-3 of human cervical tissue sections.

It is seen that the extinction coefficients for the various regions of epithelium from basal to outer surface of epithelium show variation correlated to the cellular density variations, as expected. It is worth noting that the extinction coefficient images show a better contrast compared to simple microscopic images.

| Sample category (**C**1) | Extinction Coefficient ($\mu_z$) in epithelium region. (unit:cm$^{-1}$) | | | | Extinction coefficient ratios | | |
|---|---|---|---|---|---|---|---|
| | Total Epithelium region (**C**2) | Epithelium (1/3) near basel (**C**3) | Epithelium (1/3) middle (**C**4) | Epithelium (1/3) near outer surface (**C**5) | **C**6= (**C**3/**C**4) | C7= (**C**3/**C**5) | **C**8= (**C**4/**C**5) |
| Normal | 90 | 174 | 48 | 32 | 3.63 | 5.44 | 1.50 |
| Gd-1 | 134 | 255 | 102 | 35 | 2.50 | 7.28 | 2.91 |
| Gd-2 | 297 | 506 | 322 | 54 | 1.57 | 9.37 | 5.96 |
| Gd-3 | 459 | 586 | 547 | 238 | 1.07 | 2.46 | 2.30 |

**Table 5** Average of extinction coefficients of all pixels of the epithelium of normal and various grades of dysplasia of human cervical tissue sections.

Table 5, column 2 (**C**2) shows that the average extinction coefficient of the whole epithelium region is high for higher grades of dysplastic tissue sections compared to the normal ones. It may be noted that the extinction coefficient are higher than generally reported [17]. This could be possibly be due to the samples being the dehydrated tissue sections and not soft bulk tissue. As a consequence, the morphology would be different and the scattering coefficient higher. The classification observed in column 2 (**C**2) can be further improved by taking average of the extinction coefficient over only the first 1/3$^{rd}$ region of epithelium near basal layer (shown by the arrow in figure 7). This is seen in column 3 (**C**3). Columns 4 (**C**4) and 5 (**C**5) list the average extinction coefficient for the second 1/3$^{rd}$ region of the epithelium, away from the basal layer and the last 1/3rd epithelium region, near the outer surface of epithelium respectively. Although **C**2, **C**3, **C**4 show reasonably good discrimination for grade classification with one set of samples. There may be a ±1μm thickness variation from sample to sample resulting in a variation of the average values of extinction coefficients. To take care of this, the ratio of the mean of extinction coefficients from the different regions of epithelium are shown in columns 6-8 (**C**6- **C**8). The average extinction coefficient ratio of the first 1/3rd region of epithelium including basal to middle 1/3rd region of epithelium enables us to discriminate normal and Gd-1 from Gd2 and Gd3 more consistently and independent of sample thickness. This is expected since changes in cell density are more pronounced in 2nd and 3rd regions as the grades increase. The distinction between Gd2 and Gd3 are seen in the 2nd and 3rd regions. However a consistant discrimination is noticed from the ratio of C3/C5 for all Gd2 and Gd3 samples. Since the changes between normal and Gd1 occur within the first 1/3rd of the epithelium, taking a ratio between the regions do not enhance the results.

## 4. Conclusion

Robust statistical parameters have been used to classify cervical precancerous tissue sections from their microscopic images. This method can serve as a preliminary texture analysis method prior to further image processing or even prior to other treatments for the study of tissue slides. The histopathology reports by far rely on the pathologist's judgment and thus they may differ in grading tissues that lie on the fringes of the classes. The analysis exploiting the cell density patterns of normal and the various grades of cervical dysplasia in terms of statistical parameters is seen to have the potential to be used as a quantitative parallel to the histopathology report. A few of the parameters when used in tandem result in high sensitivity and specificity for discriminating pre-cancers of the cervix.Changes in cell density along the depth of the epithelium are well reflected through the extinction coefficient images with a better contrast.

**References**

[1] Cervicalcancer.org 2010, Cervical Cancer Statistics [online] Available at: http://www.cervicalcancer.org/statistics.html [Accessed on February 2012].

[2] Information regarding various cancer statistics, their causes and risk factors may be obtained from World Health Organization [online] at http://www.who.int/mediacentre/factsheets/fs297/en/ [Accessed on February 2012].

[3] A. Wax, C. Yang, V, Backman, K. Badizadegan, C.W. Boone, R.R. Dasari and M.S. Feld, Biophysical Journal **82,** 2256–2264 ( 2002).

[4] J.R. Mourant, T.M. Johnson, S. Carpenter, A. Guerra, T. Aida and J.P. Freyer, Journal of Biomedical Optics **7(3),** 378-387(2002).

[5] J.D. Wilson, C.E. Bigelow, D.J. Calkins and T.H. Foster, Biophysical Journal **88,** 2929-2938 (2005).

[6] K Li, D. Thomasson, L. Ketai, C. Contag, M. Pomper, M. Wright and M. Bray, Clin. Infect Diseases **40(10)**, 1471-1480 (2005).

[7] K.W. Gossage, T.S. Tkaczyk, J.J. Rodriguez and J.K. Barton, Journal of Biomedical Optics **8**, 570-575 (2003).

[8] P. Bountris, E. Farantatos and N. Apostolou, in: World Academy of Science, Engineering and Technology **9**, 151-156 (2005), [online] available on: http://www.waset.org/journals/waset/v9/v9-27.pdf

[9] L.S.S. Reddy, R. Reddy, C.H. Madhu and C. Nagaraju, International Journal of Information Technology and Knowledge Management **2**, 201-204 (2010).

[10] T. Collier, M. Follen, A. Malpica and R.R. Kortum, Applied Optics **44**, 2072-2081 (2006).

[11] H.S.Sheshadri and A. Kandaswamy, Indian J Med Res, **124**, 149-154 (2006).

[12] F. Maussang, J.Chanussot and A.Hetet, in: Proc. of 7th European Conference on Underwater Acoustic, Delft, Netherlands, July 2004, pp.1133-1138.

[13] J.D.Bancroft and M. Gamble, in: J D Bancrodt (5), Theory and Practice of Histopathological Techniques, (Churchill Livingstone , China, 2002).

[14] J.A. Swets, in: Signal Detection Theory and ROC Analysis in Psychology and Diagnostics: Collected Papers, Scientific Psychology Series (Lawrence Erlbaum Associates, Mahwah, NJ, 1996).

[15] MEDCALC 2012, ROC Curve Analysis [online] Available at: http://www.medcalc.org/manual/roc-curves.php [Accessed on March 2012].

[16] V. Tuchin, in: V. Tuchin (2), Tissue Optics: Light Scattering Methods and Instruments for Medical Diagnosis,(SPIE, Washington,USA, 2007) pp.7-10.

[17] T. Collier, D. Arifler, A. Malpica, M. Follen, R. R. Kortum, IEEE Journal of Selcted Topics in Quantum Electronics **9**, 307-313 (2003).

**Authors CV:**

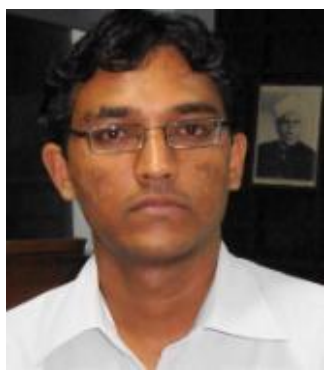

Jaidip Jagtap is currently doing his Ph.D from the Department of Physics at the Indian Institute of Technology (IIT), Kanpur, India. His areas of interest include Muller Matrix imaging, fractal analysis for 1D and 2D images, wavelet transform, medical imaging, confocal imaging for early detection of cancer.

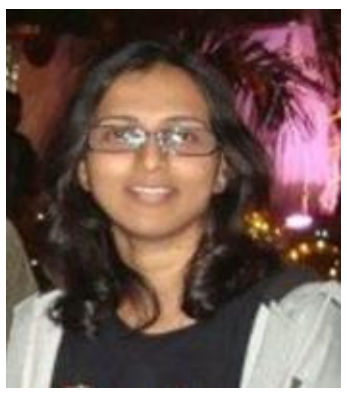

Nishigandga Patil completed her M.Tech in Center for Laser Technology from the Indian Institute of Technology (IIT), Kanpur, India in june 2012. Her areas of interest include Muller Matrix imaging, image processing, medical imaging for early detection of cancer.

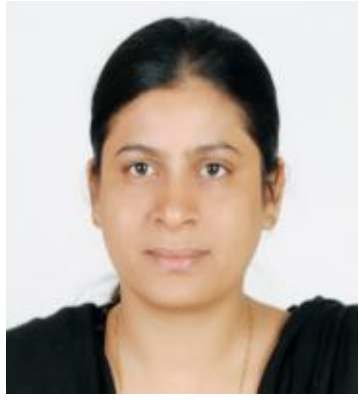

Chayanika Kala is currently working as lecturer in department of Pathology GSVM Medical College, Kanpur. She is officer in-charge cytopathology lab. She has keen interest in oncopathology, GI pathology, cytopathology and immunopathology. She has co-supervised ten post graduate thesis work and published 25 papers in various national and international journals.

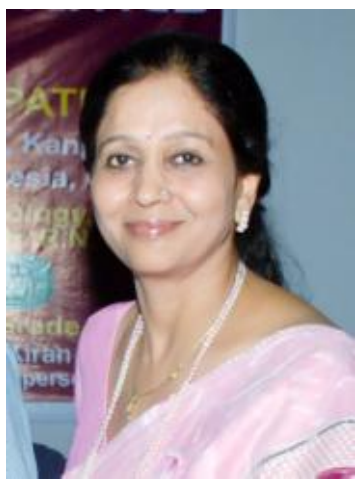 Kiran Pandey is professor and head of obstetrics and gynaecology department, G.S.V.M. medical college, kanpur (U.P). Her areas of interest include gynecological oncology, infertility obstetric, gynecological ultrasound, and gyne plastic surgery. She has 30 years experience in clinical practice and teaching and has 25 publications in national journals. She has participated and presented her work in more than 50 conferences and workshops across the country.

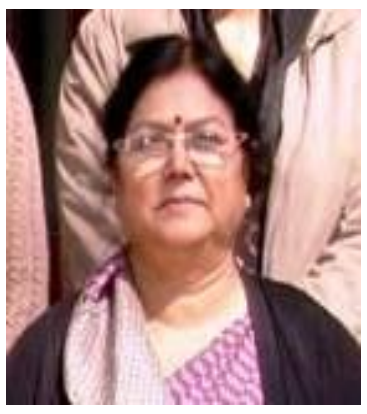 Asha Agarwal is currently working as Professor in Dept of Pathology GSVM Medical College, Kanpur. She heads the Histopathology lab and is also the nodal officer of telemedicine department. She has special interest in oncopathology and neuropathology. She has several research projects under ICMR and DMSRDE and supervised about 75 postgraduate students. She has more than 100 pubplications in reputed national and international journals.

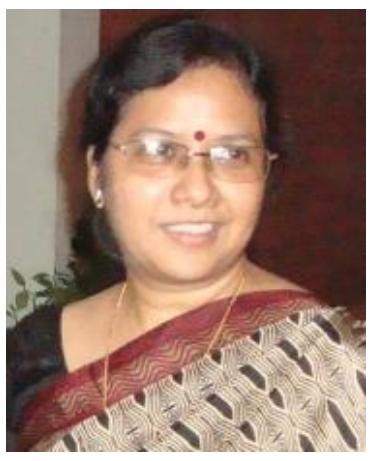 Asima Pradhan is a professor in the Department of Physics at the Indian Institute of Technology (IIT), Kanpur, India and is also a faculty member at the Center for Laser Technology. Her areas of interest include laser spectroscopy and Biophotonics. She has done her Ph.D from Institute for Ultrafast Spectroscopy and Lasers, City University of New York. Her

expertise is in fluorescence and Raman spectroscopy, light scattering spectroscopy, polarization based fluorescence and Mueller matrix imaging of bio samples. She has supervised more than 50 students towards their M.Sc, M. Tech and PhD and published more than 65 papers in various international journals. She also holds a US patent for a method to determine the tissue malignacy using time resolved fluorescence spectroscopy .